\documentclass[submission,copyright,creativecommons]{eptcs}
\providecommand{\event}{ACL2 Workshop 2017} 
\usepackage{amsmath,amssymb,amsfonts,amsthm}
\usepackage{fancyvrb}
\usepackage[shortcuts]{extdash}
\usepackage{color}
\usepackage{framed}

\usepackage{listings}
\newcommand*{\var}[1]{\mathit{#1}}

\title{Meta-extract: Using Existing Facts in Meta-reasoning}
\author{Matt Kaufmann
\institute{Department of Computer Science\\
The University of Texas at Austin\\
Austin, TX, USA}
\email{kaufmann@cs.utexas.edu}
\and
Sol Swords
\institute{Centaur Techology, Inc.\\
Austin, TX, USA}
\email{sswords@centtech.com}
}
\def\titlerunning{Meta-extract}
\def\authorrunning{M. Kaufmann \& S. Swords}
\begin{document}
\maketitle

\begin{abstract}

  ACL2 has long supported user-defined simplifiers, so-called {\em
    metafunctions} and {\em clause processors}, which are installed
  when corresponding rules of class {\tt :meta} or {\tt
    :clause-processor} are proved.  Historically, such
  simplifiers could access the logical world at execution time and
  could call certain built-in proof tools, but one could not assume the
  soundness of the proof tools or the truth of any facts extracted
  from the world or context when proving a simplifier correct.  Starting with
  ACL2 Version 6.0, released in December 2012, an additional
  capability was added which allows the correctness proofs of
  simplifiers to assume the correctness of some such proof tools and
  extracted facts.  In this paper we explain this capability and give
  examples that demonstrate its utility.

\end{abstract}

\section{Introduction}
\label{sec:intro}
The meta rule and clause processor facilities
of the ACL2 theorem proving system~\cite{acl2:home}
are designed to allow
users to write custom proof routines which, once proven
correct, can be called by the ACL2 prover\footnote{
  Clause processors may alternatively be ``trusted'' --- used without
  proof --- but at the
  cost of a {\em trust tag} marking this use as a source of potential
  unsoundness.~\cite{trusted-cl-proc}}.
These facilities descend from the meta reasoning capability
provided by earlier Boyer-Moore provers~\cite{meta}.  Relevant
background can be found in earlier ACL2 papers on meta
reasoning~\cite{meta-05} and clause processors~\cite{trusted-cl-proc},
along with ACL2 documentation~\cite{acl2:doc}.

Such meta-level proof routines have historically been limited in their use of
ACL2's database of stored facts and its built-in prover functions.
For example, a metafunction could call the ACL2 rewriter, but when
proving this metafunction correct, the result of calling the rewriter
could not be assumed to be equivalent to the input --- that is, the
rewriter could not be assumed to be correct.  Similarly,
metafunctions and clause processors could both examine the ACL2 world
(the logical state of the prover, including the database of stored
facts), but in proving the correctness of these functions, facts
extracted from the world could not be assumed to be correct.

Meta-extract is an ACL2 feature first introduced in Version 6.0
(December, 2012).  It allows for certain facts stored in the ACL2
world and certain ACL2 prover routines to be assumed
correct when proving the correctness of metafunctions and clause
processors.  In particular, additional \textit{meta-extract hypotheses}
that capture the correctness of such routines and facts are allowed to
be present in these correctness theorems. These additional hypotheses
preserve the soundness of metareasoning in ACL2 because they encode
assumptions that we are already
making --- that the facts stored in the ACL2 logical world are true
and that ACL2's prover routines are sound.

Note that meta-extract hypotheses can be used to help {\em prove}
theorems to be stored as meta rules or as clause-processor rules, but
they have no effect on how those rules are {\em applied} during
subsequent proofs.  For that, we refer readers to existing papers
\cite{meta-05,trusted-cl-proc} as well as
documentation~\cite{acl2:doc} topics including
\href{http://www.cs.utexas.edu/users/moore/acl2/manuals/current/manual/index.html?topic=ACL2\_\_\_\_META}{\underline{META}},
\href{http://www.cs.utexas.edu/users/moore/acl2/manuals/current/manual/index.html?topic=ACL2\_\_\_\_CLAUSE-PROCESSOR}{\underline{CLAUSE-PROCESSOR}},
and
\href{http://www.cs.utexas.edu/users/moore/acl2/manuals/current/manual/index.html?topic=ACL2\_\_\_\_EXTENDED-METAFUNCTIONS}{\underline{EXTENDED-METAFUNCTIONS}}.
Also see the topic
\href{http://www.cs.utexas.edu/users/moore/acl2/manuals/current/manual/index.html?topic=ACL2\_\_\_\_META-EXTRACT}{\underline{META-EXTRACT}}
for additional user-level documentation.  Note that references to
documentation topics, such as those above, are underlined to denote
hyperlinks to topics in the online
\href{http://www.cs.utexas.edu/users/moore/acl2/manuals/current/manual/index.html}{\underline{documentation}}~\cite{acl2:doc}
for ACL2 and its books.

This paper primarily addresses metafunctions rather than always
referring to both metafunctions and clause processors.  The
correctness arguments are analogous.  The meta-extract functionality
available for clause processors is a subset of the functionality available for
metafunctions: the {\em global facts} discussed below
  are available for both, but {\em contextual facts} need contexts
  that are only available below the clause level.

\subsection{Meta-extract Hypotheses}
\label{sec:intuitive}

We next introduce the various meta-extract hypotheses and
explain informally why it is sound to include them.
See the {\em Essay on Correctness of Meta Reasoning} in
the ACL2 source code for a rigorous mathematical argument.

Correctness theorems for metafunctions are stated using a
function that we will call a \textit{pseudo-evaluator}, typically
defined via the
\href{http://www.cs.utexas.edu/users/moore/acl2/manuals/current/manual/index.html?topic=ACL2\_\_\_\_DEFEVALUATOR}{\underline{\tt defevaluator}}
macro.  (Elsewhere in the ACL2
literature this is simply referred to as an evaluator; however, we
want to emphasize the difference between a pseudo-evaluator and a real
term
evaluator.)  A pseudo-evaluator is a constrained function about
which only certain facts are known; these facts are ones that would
also be true of a ``real'' evaluator capable of fully interpreting any
ACL2 term.  The intention is for these facts to suffice for
proving a metafunction correct, without the need for
a real term evaluator\footnote{
  It would be unsound to define what we are calling a real evaluator
  in ACL2 --- in particular, one that could evaluate terms containing
  calls of the evaluator itself, or functions that call the evaluator.
  In order to fix our handwavy argument, you could instead think of
  introducing an evaluator that can interpret all functions that are
  defined at the point when the metafunction is being applied.
}.
To prove a metafunction correct
the user must show that the pseudo-evaluation of the term output by
the metafunction is equal (or equivalent) to the pseudo-evaluation of
the input term.  Intuitively, if this can be proved of a
pseudo-evaluator, and the only facts known about the pseudo-evaluator
are ones that are also true of the real evaluator of ACL2 terms, then
this must also be true of the real evaluator: that is, the evaluations
of the input and output terms of the metafunction are equivalent, and
thus the metafunction is correct.

The meta-extract feature allows certain additional hypotheses in the
statement of the correctness theorem of a metafunction.  These meta-extract
hypotheses are applications of the pseudo-evaluator to calls of either the
function \texttt{meta\-/extract\-/global\-/fact+}, its less general
version \texttt{meta\-/extract\-/global\-/fact}, or the function
\texttt{meta\-/extract\-/contextual\-/fact}.
These functions produce various
sorts of terms by extracting facts from the ACL2 world and calling
ACL2 prover subroutines, constructed so that if ACL2 and its logical
state are sound, the terms produced should always be true.  For
example, these functions can produce:
\begin{itemize}
\item the body of a previously proven theorem;
\item the definitional equation of a previously defined function;
\item a term equating a call of a function on quoted constants to the
  quoted value of that call;
\item a term equating some term $a$ to the result of rewriting $a$
  using \href{http://www.cs.utexas.edu/users/moore/acl2/manuals/current/manual/index.html?topic=ACL2\_\_\_\_MFC-RW}{\underline{\tt mfc-rw}}; or
\item a term describing the type of $a$, according to ACL2's
  \href{http://www.cs.utexas.edu/users/moore/acl2/manuals/current/manual/index.html?topic=ACL2\_\_\_\_TYPE-SET}{\underline{type-set}}
  reasoning.
\end{itemize}
When we use such facts in our metafunction, meta-extract hypotheses
allow us to assume that they are true according to the
pseudo-evaluator while doing the correctness proof.

Intuitively, adding meta-extract hypotheses to a metafunction's
correctness theorem is allowable because we expect the (real)
evaluation of any term produced by one of the meta-extract functions
to return true.  If we prove the pseudo-evaluator theorem with
meta-extract hypotheses and, as before, reason that since the theorem
is true of the pseudo-evaluator, it is also true of the real
evaluator, then the final step is to observe that meta-extract
hypotheses are true using the real evaluator (or else ACL2 is already
unsound).  Therefore, even with meta-extract hypotheses, we can still
conclude that the evaluations of the metafunction's output and input
terms are equivalent.

Why not somehow axiomatize the idea that the contents of the ACL2
world are correct?  Simply put, we don't see how to do that, and ideas
we have heard along those lines have been unsound.  At the least, one
would seem to need to formalize the complex notion of a valid world.

\subsection{Organization of This Paper}

In Section~\ref{sec:meta-extract} we provide examples of meta-extract
hypotheses and summary documentation of their various forms.
Section~\ref{sec:user} presents a community book that
provides a convenient way to use the meta-extract facility.  Next we
present applications in Section~\ref{sec:applications} and finally, we
conclude with Section~\ref{sec:conclusion}.

\section{Meta-extract}
\label{sec:meta-extract}
Next we explain meta-extract by first giving two examples and then
summarizing the general forms of meta-extract hypotheses.

\subsection{Tutorial Examples}

We present two examples for the two kinds of meta-extract hypotheses,
corresponding to evaluation of calls of {\tt
  meta-extract-contextual-fact} and of {\tt meta-extract-global-fact}.
(Since a call {\tt (meta-extract-global-fact obj state)} is an
abbreviation for
{\tt (meta-extract-global-fact+ obj state state)}, we are
thus effectively illustrating {\tt meta-extract-global-fact+} as
well.)

\subsubsection{Meta-extract-contextual-fact}

Our first example is intentionally contrived and quite trivial,
intended only to provide an easy introduction to meta-extract.  It
illustrates the use of {\tt meta-extract-contextual-fact}.  The intent
is to simplify any term of the form {\tt (nth $x$ $lst$)}, when $x$ is
easily seen by ACL2 to be a symbol in the current context, to {\tt
  (car $lst$)}.

In ACL2, the use of metafunctions is always supported by an evaluator,
called a {\em pseudo-evaluator} in the preceding section.  Let us
introduce an evaluator that ``knows'' about the functions relevant to
this example.

\begin{verbatim}
(defevaluator nthmeta-ev nthmeta-ev-lst
  ((typespec-check ts x)
   (nth n x)
   (car x)))
\end{verbatim}

\noindent Next we define a metafunction, intended to replace any term
{\tt (nth n x)} by a corresponding term {\tt (car x)} when {\tt n} is
known to be a symbol using
\href{http://www.cs.utexas.edu/users/moore/acl2/manuals/current/manual/index.html?topic=ACL2\_\_\_\_TYPE-SET}{\underline{type-set}}
reasoning.\footnote{For details such as the meaning of {\tt :forcep
    nil}, see the documentation for \href{http://www.cs.utexas.edu/users/moore/acl2/manuals/current/manual/index.html?topic=ACL2\_\_\_\_META-EXTRACT}{\underline{meta-extract}}.}

\begin{verbatim}
(defun nth-symbolp-metafn (term mfc state)
  (declare (xargs :stobjs state))
  (case-match term
    (('nth n x)
     (if (equal (mfc-ts n mfc state :forcep nil)
                *ts-symbol*)
         (list 'car x)
       term))
    (& term)))
\end{verbatim}

\noindent When the input term matches {\tt (nth $n$ $x$)}, this calls
\href{http://www.cs.utexas.edu/users/moore/acl2/manuals/current/manual/index.html?topic=ACL2\_\_\_\_MFC-TS}{\underline{\tt
    mfc-ts}}
to deduce the possible types of $n$.  If that type-set equals {\tt
  *ts-symbol*} then that term must evaluate to a symbol, and the term
can be reduced to {\tt (car $x$)}.\footnote{Requiring the type-set to
  equal \texttt{*ts-symbol*} is an an unnecessarily strong check,
  chosen merely for ease of presentation: it requires that ACL2's
  determination of the term's possible types include all three of the
  basic
  types {\tt T}, {\tt NIL}, and non-Boolean symbols.}

Now we can present a meta rule with a meta-extract
hypothesis.  Without that hypothesis the formula below is
not a theorem, because the function {\tt mfc-ts} has no axiomatic
properties; all we know about it below is what we are told by the
meta-extract hypothesis, as discussed further below.

\begin{verbatim}
(defthm nth-symbolp-meta
    (implies (nthmeta-ev (meta-extract-contextual-fact `(:typeset ,(cadr term))
                                                       mfc
                                                       state)
                         a)
             (equal (nthmeta-ev term a)
                    (nthmeta-ev (nth-symbolp-metafn term mfc state) a)))
    :rule-classes ((:meta :trigger-fns (nth))))
\end{verbatim}

\noindent To see what the meta-extract hypothesis above gives us, consider the
following theorem provable by ACL2.

\begin{verbatim}
(equal (meta-extract-contextual-fact `(:typeset ,x)
                                      mfc
                                      state)
       (list 'typespec-check
             (list 'quote
                   (mfc-ts x mfc state :forcep nil))
             x))
\end{verbatim}

\noindent At a high level, this theorem shows us that {\tt
  meta-extract-contextual-fact} returns a term of the form {\tt
  (\href{http://www.cs.utexas.edu/users/moore/acl2/manuals/current/manual/index.html?topic=ACL2\_\_\_\_TYPESPEC-CHECK}{\underline{typespec-check}} (quote $ts$) x)}, which asserts that the term {\tt
  x} belongs to the set of values represented by $ts$.  The
meta-extract hypothesis applies the pseudo-evaluator to this term, and
since {\tt typespec-\allowbreak{}check} is one of its known functions, the
hypothesis reduces to
\begin{verbatim}
  (typespec-check (mfc-ts (cadr term) mfc state :forcep nil)
                  (nthmeta-ev (cadr term) a)).
\end{verbatim}
\noindent The interesting case in proving our metafunction correct is
when this {\tt mfc-ts} call equals {\tt *ts-symbol*}.  In this case the hypothesis becomes
\begin{verbatim}
  (typespec-check *ts-symbol* (nthmeta-ev (cadr term) a))
\end{verbatim}
\noindent which, when expanded, implies that the evaluation of {\tt
  (cadr term)} must be a
symbol, which enables the proof of {\tt nth-symbolp-meta}.



Recall that meta-extract hypotheses do not affect the {\em
  applications} of meta rules; they only support their proofs.
Therefore, the following test of the example above yields no surprises.

\begin{verbatim}
(defstub foo (x) t)
(thm (implies (symbolp (foo x))
              (equal (nth (foo x) y) (car y)))
     :hints (("Goal" :in-theory '(nth-symbolp-meta))))
\end{verbatim}


\subsubsection{Meta-extract-global-fact}

Our second example is from community book
  ``demos/nth-update-nth-meta-extract.lisp'', which uses {\tt
  meta-extract-global-fact}.  Let us begin by considering what problem this
book is attempting to solve.



Consider a
\href{http://www.cs.utexas.edu/users/moore/acl2/manuals/current/manual/index.html?topic=ACL2\_\_\_\_DEFSTOBJ}{\underline{\tt
    defstobj}} event, {\tt (defstobj st fld$_1$ fld$_2$ ... fld$_n$)}.
A common challenge in reasoning about stobjs is the simplification of
{\em read-over-write} terms, of the form {\tt (fld$_i$ (update-fld$_j$
  $v$ st))}, which indicate that we are to read {\tt fld$_i$} after
updating {\tt fld$_j$}.  That term simplifies to {\tt (fld$_i$ st)}
when $i \neq j$, and otherwise it simplifies to $v$.  How do we get
ACL2 to do such simplification automatically?  The following two
approaches are standard.

\begin{itemize}

\item Disable the stobj accessors and updaters after proving
  rewrite rules to simplify terms of the form {\tt (fld$_i$
    (update-fld$_j$ val st))}, to {\tt val} if $i = j$ and to {\tt
    (fld$_i$ st)} if $i \neq j$.

\item Let the stobj accessors and updaters remain enabled, relying on
  a rule such as the built-in rewrite rule {\tt nth-update-nth} to
  rewrite terms, obtained after expanding calls of the accessors and
  updaters, of the form {\tt (nth $i$ (update-nth $j$ val st))}.

\end{itemize}

\noindent The first of these requires $n^2$ rules, which is generally feasible
but can perhaps get somewhat unwieldy.  The second of these provides a
simple solution, but when proofs fail, the resulting checkpoints can
be more difficult to comprehend.

Here, we outline a solution that addresses both of these concerns: a
macro that generates a suitable meta rule.  Details of this proof
development may be found in community book
``demos/\allowbreak{}nth-update-nth-meta-extract.lisp''.
First we introduce our metafunction.  Next, we prove a meta rule for a
specific stobj.  Finally, we mention a macro that generates a version
of this rule that is suitable for an arbitrary specified stobj.

Our metafunction returns the input term unchanged unless it is of the
form {\tt ($r$ ($w$ $v$ $x$))}, where $r$ and $w$ are {\em reader}
(accessor) and {\em writer} (updater) functions defined to be calls of
{\tt nth} and {\tt update-nth}, on explicit indices: {\tt ($r$ $x$)}
$=$ {\tt (nth '$i$ $x$)} and {\tt ($w$ $v$ $x$)} $=$ {\tt (update-nth
  '$i$ $v$ $x$)}.  In that case, the function {\tt
  nth-update-nth-meta-fn-new-term} computes a new term: $v$ if $i =
j$, and otherwise, {\tt ($r$ $x$)}.

\begin{verbatim}
(defun nth-update-nth-meta-fn (term mfc state)
  (declare (xargs :stobjs state)
           (ignore mfc))
  (or (nth-update-nth-meta-fn-new-term term state)
      term))
\end{verbatim}

Notice below that in computing the new term, the definitions of the
reader and writer are extracted from the logical world using the
function, {\tt meta-extract-formula}, which returns the function's
definitional equation.  For example:

\begin{verbatim}
ACL2 !>(meta-extract-formula 'atom state)
(EQUAL (ATOM X) (NOT (CONSP X)))
ACL2 !>
\end{verbatim}

\noindent We thus rely on the correctness of {\tt
  meta-extract-formula} for the equality of the input term and the
term returned by the following function.

\begin{Verbatim}[commandchars=\\\{\},fontsize=\small]
(defun nth-update-nth-meta-fn-new-term (term state)
  (declare (xargs :stobjs state))
  (case-match term
    ((reader (writer val x))
     (and (not (eq reader 'quote))
          (not (eq writer 'quote))
          (let* ((reader-formula (and (symbolp reader)
                                      (meta-extract-formula reader state)))
                 (i-rd (fn-nth-index reader reader-formula)))
            (and
             i-rd {\em ; the body of reader is (nth 'i-rd ...)}
             (let* ((writer-formula (and (symbolp writer)
                                         (meta-extract-formula writer state)))
                    (i-wr (fn-update-nth-index writer writer-formula)))
               (and
                i-wr {\em ; the body of writer is (update-nth 'i-wr ...)}
                (if (eql i-rd i-wr)
                    val
                  (list reader x))))))))
    (& nil)))
\end{Verbatim}

Next we introduce a (pseudo-)evaluator to use in our meta rule.

\begin{verbatim}
(defevaluator nth-update-nth-ev nth-update-nth-ev-lst
  ((nth n x)
   (update-nth n val x)
   (equal x y)))
\end{verbatim}

We need to define one more function before presenting our meta rule.
It takes as input a term {\tt ($f$ $t_1$ $\ldots$ $t_n$)} with list of
formals {\tt ($v_1$ $\ldots$ $v_n$)}, and builds an alist that maps
each $v_i$ to the value of $t_i$ in a given alist.  Like our
metafunction, it consults {\tt meta-extract-formula} to obtain the
formal parameters of $f$.  An earlier attempt to use the function,
{\tt formals}, failed: the meta rule's proof needs these formals to
connect to those found by our metafunction.

\begin{Verbatim}[commandchars=\\\{\},fontsize=\small]
(defun meta-extract-alist-rec (formals actuals a)
  (cond ((endp formals) nil)
        (t (acons (car formals)
                  (nth-update-nth-ev (car actuals) a)
                  (meta-extract-alist-rec (cdr formals) (cdr actuals) a)))))

(defun meta-extract-alist (term a state)
  (declare (xargs :stobjs state :verify-guards nil))
  (let* ((fn (car term))
         (actuals (cdr term))
         (formula (meta-extract-formula fn state)) {\em ; (equal (fn ...) ...)}
         (formals (cdr (cadr formula))))
    (meta-extract-alist-rec formals actuals a)))
\end{Verbatim}

\noindent We now define a stobj and prove a corresponding meta rule.

\begin{Verbatim}[commandchars=\\\{\},fontsize=\small]
(defstobj st
  fld1 fld2 fld3 fld4 fld5 fld6 fld7 fld8 fld9 fld10
  fld11 fld12 fld13 fld14 fld15 fld16 fld17 fld18 fld19 fld20)

(defthm nth-update-nth-meta-rule-st
  (implies
   (and (nth-update-nth-ev {\em ; (f (update-g val st))}
         (meta-extract-global-fact (list :formula (car term)) state)
         (meta-extract-alist term a state))
        (nth-update-nth-ev {\em ; (update-g val st)}
         (meta-extract-global-fact (list :formula (car (cadr term)))
                                   state)
         (meta-extract-alist (cadr term) a state))
        (nth-update-nth-ev {\em ; (f st) -- note st is (caddr (cadr term))}
         (meta-extract-global-fact (list :formula (car term)) state)
         (meta-extract-alist (list (car term)
                                   (caddr (cadr term)))
                             a state)))
   (equal (nth-update-nth-ev term a)
          (nth-update-nth-ev (nth-update-nth-meta-fn term mfc state) a)))
  :hints ...
  :rule-classes ((:meta :trigger-fns (fld1 fld2 ... fld20))))
\end{Verbatim}

\noindent The proof of the theorem below takes no measurable time, and
applies the metafunction proved correct above.

\begin{verbatim}
   (in-theory (disable fld1 ... fld20 update-fld1 ... update-fld20))

   (defthm test1
     (equal (fld3 (update-fld1 1
                   (update-fld2 2
                    (update-fld3 3
                     (update-fld4 4
                      (update-fld3 5
                       (update-fld6 6 st)))))))
            3))
\end{verbatim}

Notice that there is nothing about the meta rule above that is
specific to the particular stobj, {\tt st}, except for the {\tt
  :trigger-fns} that it specifies.  In the community book mentioned
above (``demos/nth-\allowbreak{}update-\allowbreak{}nth-\allowbreak{}meta-\allowbreak{}extract.lisp''), we define a macro
that automates the generation of such a meta rule for an
arbitrary stobj.  Our macro takes the name of a stobj, $s$, and does two
things: it disables all of the stobj's accessors and updaters, and it
proves a meta rule that simplifies every term of the form {\tt ($r$
  ($w$ $v$ $s$))}, where $r$ and $w$ are an accessor and updater,
respectively, for the stobj $s$.

The meta rule above uses three \texttt{meta-extract-global-fact}
hypotheses corresponding to uses of three particular facts.
Enumerating all of the facts potentially used in the metafunction in
this way could easily become unwieldy; in Section \ref{sec:user} we
discuss a utility that solves this problem, allowing one
\texttt{meta-extract-global-fact} hypothesis and one
\texttt{meta-extract-contextual-fact} hypothesis to cover all of the
facts that might be used.

\subsection{General Forms}
\label{sec:general}

In this section we summarize briefly the forms of meta-extract
hypotheses.  Additional details may be found in the documentation for
\href{http://www.cs.utexas.edu/users/moore/acl2/manuals/current/manual/index.html?topic=ACL2\_\_\_\_META-EXTRACT}{\underline{meta-extract}}.

Below, let {\tt evl} be the pseudo-evaluator (see
Section~\ref{sec:intuitive}) used in a meta rule.  The two primary
forms of meta-extract hypotheses that can be used in a meta rule are
as follows.  In the first, {\tt aa} represents an arbitrary term; in
the second, {\tt a} must be the second argument of the two calls of
the pseudo-evaluator in the conclusion of the theorem.

\begin{verbatim}
    (evl (meta-extract-global-fact obj state) aa)
    (evl (meta-extract-contextual-fact obj mfc state) a)
\end{verbatim}

The second form above is only legal for meta rules about extended
metafunctions (which take arguments {\tt mfc} and {\tt state}).  The
first form above is actually equivalent to the first form below, which
in turn is a special case of the second form below.

\begin{verbatim}
    (evl (meta-extract-global-fact+ obj state state) aa)
    (evl (meta-extract-global-fact+ obj st state) aa)
\end{verbatim}

The last form supports clause processors that modify state as long as
they do not change the logical world; it produces the same value as
the previous form as long as both {\tt st} and {\tt state} have equal
world fields.

If the arguments to the meta-extract-* function are somehow malformed,
then it returns the trivial term {\tt 'T}, which is not of any use in
proving a metafunction correct.  Otherwise, each invocation produces a
term that states the ``correctness'' of an invocation of some
utility, such as {\tt mfc-rw+} or {\tt meta-\allowbreak{}extract-\allowbreak{}formula}.  The
terms produced for different kinds of invocation vary according to the
particular concept of correctness appropriate to the utility in
question.


We now describe the allowed {\tt obj} arguments to {\tt
  meta-extract-global-fact} and the terms they produce.  The
documentation for
\href{http://www.cs.utexas.edu/users/moore/acl2/manuals/current/manual/index.html?topic=ACL2\_\_\_\_EXTENDED-METAFUNCTIONS}{\underline{extended-metafunctions}}
explains meta-level functions discussed below, such as {\tt mfc-rw}
and {\tt mfc-ap}.

\begin{itemize}

\item {\tt (:formula $f$)} produces the value of {\tt
    (meta-extract-formula $f$ state)}, which allows metafunctions to
  assume that any invocation of {\tt meta-extract-formula} produces a
  true formula.  {\tt Meta-extract-formula} looks up various kinds of
  formulas from the world:
  \begin{itemize}
  \item the body of $f$ if it is a theorem name
  \item the definitional equation of $f$ if it is a defined function
  \item the constraint of $f$ if it is a constrained function
  \item the defchoose axiom of $f$ if it is a \href{http://www.cs.utexas.edu/users/moore/acl2/manuals/current/manual/index.html?topic=ACL2\_\_\_\_DEFCHOOSE}{\underline{\tt defchoose}} function.
  \end{itemize}


\item {\tt (:lemma $f$ $n$)} produces a term corresponding to the
  $n$th rewrite rule stored in the {\tt lemmas} property of function
  $f$, which allows any such rule to be assumed correct in a
  metafunction.  The term returned is the value of
  \begin{lstlisting}[basicstyle=\linespread{0.4}\normalsize\ttfamily]
    (rewrite-rule-term (nth '<$n$> (getpropc '<$f$> 'lemmas nil (w state))))<\textrm{,}>
  \end{lstlisting}
  assuming that $n$ is a valid index (does not exceed
  the length of the indicated list).  Here {\tt rewrite-\allowbreak{}rule-\allowbreak{}term}
  transforms a {\tt rewrite-rule} record structure into a term such as
  \begin{lstlisting}[basicstyle=\linespread{0.4}\normalsize\ttfamily]
    (implies <$\var{hyps}$> (<$\var{equiv}$> <$\var{lhs}$> <$\var{rhs}$>))<\textrm{.}>
  \end{lstlisting}
  The rewrite rules stored in {\tt :lemmas} are not simply theorem
  bodies, which could be accessed using {\tt meta-extract-formula},
  but rewrite rule structures, which separately store the left-hand
  side, right-hand side, equivalence relation, and hypotheses, and
  also contain heuristic information such as the backchain limits and
  match-free mode.

\item {\tt (:fncall $f$ $L$)} produces a term {\tt (equal $c$ '$v$)}, where
    $c$ is the call that applies $f$ to the quotations of the values in argument list $L$, and $v$ is
    the value of that call computed by
    \href{http://www.cs.utexas.edu/users/moore/acl2/manuals/current/manual/index.html?topic=ACL2\_\_\_\_MAGIC-EV-FNCALL}{\underline{\tt
        magic-\allowbreak{}ev-\allowbreak{}fncall}}.  This allows
    metafunctions to assume that {\tt magic-ev-fncall}
    correctly evaluates function applications.

\end{itemize}

The allowed {\tt obj} arguments to {\tt meta-extract-contextual-fact} are as follows.

\begin{itemize}

\item {\tt (:typeset $\var{term}$)} produces a term stating the correctness of
    the type-set produced by {\tt mfc-ts} for $\var{term}$.  Specifically, it produces the term
    {\tt (typespec-check '$\var{ts}$ $\var{term}$)}, where $\var{ts}$ is
    the result of {\tt (mfc-ts $\var{term}$ mfc state :forcep nil :ttreep nil)} and
    {\tt (typespec-check ts val)} is true when {\tt val}'s actual type is in the type-set {\tt ts}.

  \item {\tt (:rw+ $\var{term}\ \var{alist}\ \var{obj}\ \var{equiv}$)}
    produces a term stating that {\tt mfc-rw+} correctly rewrites
    $\var{term}$ under substitution $\var{alist}$ with objective
    $\var{obj}$ under equivalence relation $\var{equiv}$.  The form of the term produced is
  \begin{lstlisting}[basicstyle=\linespread{0.4}\normalsize\ttfamily]
    (<$\var{equiv}$> <$\var{term}'$> <$\var{rw}$>)
  \end{lstlisting}
  where $\var{term}'$ is the new term formed by substituting
  $\var{alist}$ into $\var{term}$ and $\var{rw}$ is the result of the call
  {\tt (mfc-rw+ $\var{term}\ \var{alist}\ \var{obj}\ \var{equiv}$ mfc state :forcep nil :ttreep nil)}.
  The $\var{equiv}$ argument may also be {\tt T}, meaning
  {\tt IFF}, or {\tt NIL}, meaning {\tt EQUAL}.

\item {\tt (:rw $\var{term}\ \var{obj}\ \var{equiv}$)} is similar to the
  {\tt :rw+} form above, but instead of {\tt mfc-rw+} it supports {\tt
    mfc-rw}, which takes no $\var{alist}$ argument.  Instead, {\tt
    NIL} is used for the substitution.

\item {\tt (:ap $\var{term}$)} uses {\tt mfc-ap} to derive a linear
  arithmetic contradiction indicating that $\var{term}$ is false, and
  produces {\tt (not $\var{term}$)} if that is successful, that is, if
  {\tt (mfc-ap $\var{term}$ mfc state :forcep nil)} returns true;
  otherwise it just produces {\tt 'T}.

\item {\tt (:relieve-hyp $\var{hyp}\ \var{alist}\ \var{rune}\
    \var{target}\ \var{backptr}$)} uses {\tt mfc-relieve-hyp} to
  attempt to prove that $\var{hyp}$ holds under substitution
  $\var{alist}$, and produces the substitution of
  $\var{alist}$ into $\var{hyp}$ if successful, that is, if
  {\tt (mfc-relieve-hyp $\var{hyp}\ \var{alist}\ \var{rune}\
    \var{target}\ \var{backptr}$ mfc state :forcep nil :ttreep nil)}
  returns true; otherwise, {\tt 'T}.

\end{itemize}


We have seen there are two forms of meta-extract hypotheses: {\tt
  meta-extract-contextual-fact} and {\tt meta-extract-global-fact+}
(and its less general form, {\tt meta-extract-global-fact}).  We could
have split each of these into several forms, resulting in eight forms
for the eight kinds of values of {\tt obj} listed above.  However, in
the next section we describe a utility that essentially generalizes
the two supported forms, which eliminates the need for the user to
think about the precise values of {\tt obj}; it was convenient to
generalize two supported forms rather than eight.

\section{Using ``meta-extract-user.lisp''}
\label{sec:user}
The community book ``clause-processors/meta-extract-user.lisp'' is
designed to allow more convenient use of the meta-extract facility.
The main contribution of this book is in the event-generating macro
\href{http://www.cs.utexas.edu/users/moore/acl2/manuals/current/manual/index.html?topic=ACL2\_\_\_\_DEF-META-EXTRACT}{\underline{\tt def-meta-extract}}. For a given pseudo-evaluator \texttt{evl},
\texttt{def\-/meta\-/extract} produces macros
\texttt{evl\-/meta\-/extract\-/contextual\-/facts} and
\texttt{evl\-/meta\-/extract\-/global\-/facts} that expand to meta-extract
hypotheses where the \texttt{obj} argument is a call of a ``bad-guy'' (Skolem)
function.  This essentially universally quantifies the \texttt{obj}
argument: informally, the term
\begin{lstlisting}
  <$\var{obj}_0 = $> (evl-meta-extract-contextual-bad-guy a mfc state)
\end{lstlisting}
is
some {\tt obj} for which the formula
\begin{lstlisting}
  <$\varphi = $> (evl (meta-extract-contextual-fact obj mfc state) a)
\end{lstlisting}
is false, if any such {\tt obj} exists; so by
asserting that $\varphi$ is true for $\var{obj}_0$, we assert $(\forall\ \var{obj})\ 
\varphi$.  Therefore,
\begin{verbatim}
  (evl (meta-extract-contextual-fact
         (evl-meta-extract-contextual-bad-guy a mfc state)
         mfc state)
       a)
\end{verbatim}
implies that for any \texttt{obj},
\begin{verbatim}
  (evl (meta-extract-contextual-fact obj mfc state) a).
\end{verbatim}

The \texttt{def\-/meta\-/extract} utility also proves several theorems
about the pseudo-evaluator.  They are proven by
functional instantiation of similar theorems proved in
``meta-extract-user.lisp'' about a base evaluator that only supports
the six functions {\tt typespec-check}, {\tt if}, {\tt equal}, {\tt
  not}, {\tt iff}, and {\tt implies}; this means that {\tt evl} must
also support at least these six functions for the utility to work.

The theorems proved by \texttt{def\-/meta\-/extract} obviate the need for the user to reason about
the specifics of the definitions of
\texttt{meta\-/extract\-/contextual\-/fact} and
\texttt{meta\-/extract\-/global\-/fact+} and the proper construction
of their \texttt{obj} arguments, while still supporting all the facilities listed in Section~\ref{sec:general}.
For example, this rule shows that
\texttt{(evl\-/meta\-/extract\-/global\-/facts)} implies the correctness of \texttt{meta\-/extract\-/formula} (and makes no reference to the form of the \texttt{obj} argument to \texttt{meta\-/extract\-/global\-/fact}):
\begin{verbatim}
 (defthm evl-meta-extract-formula
   (implies (and (evl-meta-extract-ev-global-facts)
                 (equal (w st) (w state)))
            (evl (meta-extract-formula name st) a)))

\end{verbatim}
This rule shows that
\texttt{(evl\-/meta\-/extract\-/contextual\-/facts a)} implies the correctness
of \texttt{mfc-rw+} (specifically, when \texttt{nil}, meaning
\texttt{equal}, is given as the equivalence relation argument):
\begin{verbatim}
 (defthm evl-meta-extract-rw+-equal
    (implies (evl-meta-extract-contextual-facts a)
             (equal (evl (mfc-rw+
                          term alist obj nil
                          mfc state :forcep nil)
                         a)
                    (evl (sublis-var alist term) a))))
\end{verbatim}
The above rule involves the system function {\tt sublis-var}, which
substitutes {\tt alist} into {\tt term} but reduces ground calls of
primitive functions\footnote{By ``primitive functions'' we mean
  built-in functions such as {\tt binary-+} that do not have defining
  events --- that is, those found in the ACL2 constant {\tt
    *primitive-formals-and-guards*}.} to their values.  In order to
reason about {\tt sublis-var}, the pseudo-evaluator should support all of
the primitive functions that it treats specially.  The community book
``clause-processors/sublis-var-meaning.lisp'' defines a
pseudo-evaluator {\tt cterm-ev} that supports exactly these functions
and proves the following theorem that describes how {\tt sublis-var}
evaluates:
\begin{verbatim}
(defthm eval-of-sublis-var
  (implies (and (pseudo-termp x)
                (not (assoc nil alist)))
           (equal (cterm-ev (sublis-var alist x) a)
                  (cterm-ev x (append (cterm-ev-alist alist a) a)))))
\end{verbatim}

To reason about {\tt mfc-rw+}, {\tt mfc-rw}, and {\tt
  mfc-relieve-hyp}, whose meta-extract assumptions all involve {\tt
  sublis-var}, one can define a pseudo-evaluator that supports both
the functions required for \texttt{def\-/meta\-/extract} and for {\tt
  sublis-var}.  Community book ``centaur/misc/context-rw.lisp'', for
example, defines a pseudo-evaluator supporting all these functions,
uses {\tt def-meta-extract}, and functionally instantiates the above
theorem about {\tt sublis-var} to allow it to reason about {\tt
  mfc-rw+}.






\section{Applications}
\label{sec:applications}

\subsection{GL Symbolic Interpreter}

The
\href{http://www.cs.utexas.edu/users/moore/acl2/manuals/current/manual/index.html?topic=ACL2\_\_\_\_GL}{\underline{GL}}
framework for bit-level symbolic execution~\cite{gl-diss,
  bit-blasting-GL} is an important tool for hardware verification
efforts at Centaur Technology \cite{centaur-framework}.  To prove a
theorem, GL attempts to symbolically interpret the conclusion and
render it into a Boolean formula which can be proved via a
satisfiability check.  Given a bit-level representation of each
variable, the symbolic interpreter recursively computes a bit-level representation
of each subterm.  It expands function definitions as needed, down to
certain primitive functions for which support is built in.  Recent
versions can also apply rewrite rules.  It is implemented as a clause
processor and uses meta-extract to look up function definitions and
rewrite rules from the world and to evaluate ground terms using
\texttt{magic-ev-fncall}.

GL was originally written before meta-extract, and used two tricks to
replace its functionality.  The complexity of these tricks reflects
the utility of meta-extract, since it allows GL to now avoid these
desperate measures.
\begin{itemize}
\item In order to justify the correctness of function definitions, GL
  would keep track of all definitions that were used, and return each
  definitional equation as an additional proof obligation from the
  clause processor.  GL proof hints were orchestrated so as to use
  \texttt{:by} hints to attempt to cheaply discharge these
  obligations.  GL did not yet use rewrite rules, but they could have
  been handled similarly.
\item In order to apply functions to concrete arguments, each GL
  clause processor had an apply function that could call a fixed set
  of functions by name using a \texttt{case} statement.  GL provided
  automation for defining new such clause processors so that users
  could build in their own set of functions.
\end{itemize}

\subsection{Rewrite-bounds}

The community book ``centaur/misc/bound-rewriter.lisp'' provides a
tool for solving certain inequalities: it replaces subterms of an
inequality with known bounds if those subterms are in monotonic
positions.  For example, the term $a-b$ monotonically decreases as $b$
increases, so if we wish to prove $c<a-b$ and we know $B \geq b$, then
it suffices to prove $c<a-B$.  While this example would be easily
handled by ACL2's linear arithmetic solver, there are problems that
the bound rewriter can handle easily that overwhelm ACL2's nonlinear
solver and cannot be solved with linear arithmetic -- e.g.,
\begin{verbatim}
(implies (and (rationalp a) (rationalp b) (rationalp c)
              (<= 0 a) (<= 0 b) (<= 1 c)
              (<= a 10) (<= b 20) (<= c 30))
         (<= (+ (* a b c) (* a b) (* b c) (* a c))
             (+ (* 10 20 30) (* 10 20) (* 20 30) (* 10 30))))
\end{verbatim}
This formula can't be solved with linear arithmetic, because it is not a
linear problem.  (If each product of variables is replaced by a fresh
variable, the result is clearly not a theorem.)
Moreover, the hint \texttt{:nonlinearp t} causes ACL2 to hang indefinitely.
However, the hint
\begin{verbatim}
(rewrite-bounds ((<= a 10)
                 (<= b 20)
                 (<= c 30)))
\end{verbatim}
solves it instantaneously by replacing upper-boundable occurrences of
\texttt{a} by 10, \texttt{b} by 20, and \texttt{c} by 30.  The same
results are obtained --- a quick proof using {\tt rewrite-bounds} but
an indefinite hang using nonlinear arithmetic --- if the arithmetic
expression on the last line of the theorem is replaced by its value,
7100.

The bound rewriter tool is implemented as a meta rule and uses
meta-extract extensively.  To determine which subterms are in
monotonic positions, it uses type-set reasoning to determine the signs
of subterms.  For example, $a \cdot b$ increases as $b$ increases if
$a$ is nonnegative and decreases as $b$ increases if $a$ is
nonpositive; if we can't (weakly) determine the sign of $a$, then we
can't replace $b$ or any subterm with a bound.  To determine whether a
proposed bound of a subterm is (contextually) true, it uses
\texttt{mfc-relieve-hyp} to show it by rewriting, and if that fails,
\texttt{mfc-ap} to show it by linear arithmetic reasoning.  The
correctness of these uses of ACL2 reasoning utilities are justified by
meta-extract-contextual-fact hypotheses.

\subsection{Context-rw}

A meta rule for context-sensitive rewriting, accomplishing something
similar to Greve's ``Nary'' framework \cite{greve06}, is defined in
``centaur/misc/context-rw.lisp'' (see
\href{http://www.cs.utexas.edu/users/moore/acl2/manuals/current/manual/index.html?topic=ACL2\_\_\_\_CONTEXTUAL-REWRITING}{\underline{contextual-rewriting}}).
Like Nary, it supports an analogue of congruence reasoning using
contexts that are more general than equivalence relations.  An example
of its use is to allow, e.g.,
\begin{verbatim}
(logbitp 4 (logand (logior a b c (logapp 6 d e)) f g))
\end{verbatim}
\noindent to be simplified to
\begin{verbatim}
(logbitp 4 (logand (logior a b c d) f g))
\end{verbatim}
without defining a rewrite rule specifically for that case. The
context rewriter uses certain theorems to justify rewriting subterms
under new contexts, that is, with new calls wrapped around them.  In
this case we might have a rule that says that \texttt{(logbitp 4 ...)}
induces a \texttt{(logand 16 ...)} context; this context can then be
propagated through the \texttt{logand} and \texttt{logior} down to the
\texttt{logapp} call, which simplifies under that context to just
\texttt{d}, using a traditional rewrite rule such as the following:
\begin{verbatim}
(implies (and (syntaxp (and (quotep n) (quotep m)))
              (equal (logtail m n) 0))
         (equal (logand n (logapp m a b))
                (logand n a)))
\end{verbatim}
\noindent When interpreted as a context rule, the following says that
\texttt{logbitp} induces a \texttt{logand} context, by directing
replacement of an instance of the right-hand side of the equality by
the corresponding instance of the left-hand side.
\begin{verbatim}
(implies (syntaxp (quotep n))
         (equal (logbitp n (logand (ash 1 (nfix n)) m))
                (logbitp n m)))
\end{verbatim}
\noindent And the following rule says that \texttt{logior} propagates such a
\texttt{logand} context onto its second argument:
\begin{verbatim}
(implies (syntaxp (quotep n))
         (equal (logand n (logior a (logand n b)))
                (logand n (logior a b))))
\end{verbatim}
\noindent A syntactic requirement for context rules is that the left-
and right-hand sides must be identical except that some subterm of the
LHS is replaced by a variable in the RHS (in this case \texttt{b}),
and that is the only occurrence of that variable in the RHS.  That
variable corresponds to the subterm onto which the context will be
propagated, and the corresponding subterm of the LHS reflects the
context that will be propagated onto it.  So the rule immediately above says:
\begin{quote}
When \texttt{logior} occurs under a \texttt{(logand n ...)} context where
\texttt{n} is a constant, propagate the \texttt{(logand n ...)} context
onto the second argument of the \texttt{logior}.
\end{quote}
\noindent Experienced ACL2 users might note that the roles of the
left- and right-hand sides are essentially reversed in this usage;
this is because often the reverse of a good context-propagation rule
is also a good rewrite rule.

The context rewriter uses meta-extract in order to trust rewrite rules
extracted from the world and results from \texttt{mfc-rw+} and
\texttt{mfc-relieve-hyp}, which it uses, respectively, to simplify
subterms under new contexts and to discharge hypotheses necessary for
applying the context rules.

\subsection{Others}

A few other utilities from the community books use
meta-extract solely to be able to extract a formula from the world
(using \texttt{meta-extract-formula}) and assume it to be true.  For
example, ``clause-processors/witness-cp.lisp'' provides a framework
for reasoning about quantification (see
\href{http://www.cs.utexas.edu/users/moore/acl2/manuals/current/manual/index.html?topic=ACL2\_\_\_\_WITNESS-CP}{\underline{witness-cp}});
it uses \texttt{meta-extract-formula} to look up a stored fact showing
that a term representing a universal quantification implies any
instance of the quantified formula.  A second,
``clause-processors/just-expand.lisp'', provides a clause processor
and meta rule that force expansion of certain terms, somewhat similar
to the \texttt{:expand} hint.  A third,
``clause-processors/replace-equalities.lisp'', provides a tool for
replacing known equalities in ways that the
rewriter can't, e.g., replacing a variable with a term.  For example, the
following is not a valid rewrite rule because its left-hand side is a
variable, but it could be a good replace-equalities rule:
\begin{verbatim}
 (implies (matches pattern x)
          (equal x
                 (patternsubst pattern (sigma pattern x))))
\end{verbatim}




\section{Conclusion}
\label{sec:conclusion}

This paper explains meta-extract hypotheses and shows how they can be
put to good use, either directly or by way of the {\tt
def-meta-extract} utility to obtain two simple, general meta-extract
hypotheses.  The implementation and logical justification for
meta-extract are delicate, so it made sense to implement only the
primitive notions of Section~\ref{sec:meta-extract} in ACL2 and then
introduce {\tt def-meta-extract} in a book, with less trusted code as
a result.

The meta-extract facility has been successfully used to create
specialized proof tools that are admitted by ACL2 as fully verified
metafunctions or clause-processors, including the GL bit-blasting
framework, which is in daily use at Centaur Technology.  We hope that
this paper contributes to wider successful use of the meta-extract
feature.

\section*{Acknowledgments}

We thank J Moore for helpful discussions.
We also thank the referees for useful feedback.
This material is based upon work supported in part by DARPA under
Contract No. FA8750-15-C-0007 (subcontract 15-C-0007-UT-Austin)
and by ForrestHunt, Inc.

\bibliographystyle{eptcs}
\bibliography{paper}
\end{document}